\documentclass[twocolumn,prc,showpacs,a4]{revtex4}
\usepackage{psfig,epsfig,graphicx,float,pst-all}
\usepackage{amssymb}
\renewcommand{\=}{~=~}
\newcommand{\be}{\begin{equation}}
\newcommand{\ee}{\end{equation}}
\newcommand{\bc}{\begin{center}}
\newcommand{\ec}{\end{center}}

\begin{document}

\title{Soft triaxial roto-vibrational motion in the vicinity of $\gamma=\pi/6$}
\author{Lorenzo Fortunato \footnote{E-mail: {\tt fortunat@pd.infn.it} }}
\affiliation{Vakgroep subatomaire en stralingsfysica, Proeftuinstraat 86,
B-9000, Ghent (Belgium)}

\begin{abstract}
A solution of the Bohr collective hamiltonian for the $\beta-$soft, 
$\gamma-$soft triaxial rotor with $\gamma \sim \pi/6$ is presented 
making use of a harmonic potential in $\gamma$ and  Coulomb-like and 
Kratzer-like potentials in $\beta$. 
It is shown that, while the $\gamma-$angular part in the present case 
gives rise to a straightforward extension of the rigid triaxial rotor 
energy in which an additive harmonic term appears, the inclusion of the 
$\beta$ part results instead in a non-trivial expression for the spectrum. 
The negative anharmonicities of the energy levels with respect to a 
simple rigid model are in qualitative agreement with general trends in
the experimental data.
\end{abstract}
\maketitle

The search for new solutions of the Bohr 
collective hamiltonian has recently experienced a resurgence, 
mainly due to influential studies upon shape phase transitions 
\cite{Iac,Iac2}. 
The occurrence of analytically solvable models has served to 
define benchmarks for the analysis of experimental 
nuclear spectra. As long as the quadrupole degree of freedom is considered,
the nuclear Bohr-Mottelson collective model has a few well-known complete
solutions that correspond to spherical or axially symmetric equilibrium 
intrinsic shapes. In some case, when the full solution is a demanding
task, one has usually resorted to the so-called {\it rigid} models in which
one or both of the deformation variables, $\beta$ and $\gamma$, is 
constrained to take a fixed value. The opposite situation, in which either 
the $\beta$ or $\gamma$ variable (or both) is not forced to take a fixed 
value, is called {\it soft}.

The purpose of the present paper is to present a novel
solution of the Bohr collective hamiltonian for the $\beta-$soft, 
$\gamma-$soft triaxial rotor with $\gamma$ around $\pi/6$.
Our model include $\beta$ and $\gamma$ vibrations as well as rotations 
about a non-axially symmetric equilibrium intrinsic shape.

Years ago Davydov \cite{Dav} proposed a model for a rigid triaxial rotor, 
described diagonalizing the
hamiltonian for a rigid rotor with three different moments of inertia. 
The corresponding soft cases, that are described 
by the full Bohr hamiltonian with some $V(\beta,\gamma)$ potential 
(that displays a pocket whose minimum lies in $\beta_0\ne 0,\gamma_0\ne 0$), 
has been only partially or numerically treated (see also \cite{Mey}).
Davydov discussed also a model with soft behaviour with respect to $\beta$
\cite{DavCh} and a model with soft behaviour with respect to $\gamma$
\cite{Dav2}. These approaches give rather complicated expressions for the 
spectrum. The latter, in which the effective potential in $\gamma$ is 
expanded in a series and truncated neglecting terms of order higher than 
the second, has some similarity with the present approach.
Our model, although restricted to $\gamma \sim \pi/6$, is easier to handle 
and gives in a straightforward fashion the full energy spectrum, while 
the cited Davydov's treatments give only the energies of a limited number 
of states.

This study is partly born out from the necessity to overcome the 
reservations on the above models (see for example \cite{Rowe}) about the 
neglected role of the fluctuations of the dynamic moments of inertia 
and partly to show that the na\"{\i}ve idea that the spectrum of a soft
triaxial rotor will be the sum of a rigid rotor part plus a harmonic
$\beta-$ and harmonic $\gamma-$vibrations is essentially incorrect. 
In the present case the solution of the $\gamma-$angular part is indeed 
a straightforward 
extension of the rigid triaxial rotor energy in which an additive harmonic 
term appears, but the inclusion of the $\beta$ part gives rise to a peculiar
combination of the various terms.

A closely connected solution of the Bohr hamiltonian has recently 
appeared as a preprint \cite{Bon}. The common point between this work
and ours is the presence of a minimum in $\gamma=30^o$.
In the cited work a critical point symmetry, called Z(5), is introduced
for the prolate to oblate shape phase transition. The authors 
discuss the solution and compare their predictions with experimental 
data in the Pt region, obtaining a good agreement. The main difference 
between the two approaches lies essentially in the expressions of the 
potentials. Their potential may be 
written as $u_1(\beta)+u_2(\gamma)$, where a harmonic oscillator 
is used for the $\gamma$ variable, and where an infinite square
well is used for the $\beta$ potential, similarly to what has been done
in Refs. \cite{Iac,Iac2} for the so-called E(5) and X(5) symmetries, 
thus achieving 
an approximate separation of variables. Anticipating the following
discussion we mention that in the present solution a potential
of the form (\ref{pote}) makes exact the separation of variables and
Coulomb and Kratzer potentials are used in the $\beta$ variable.

Two very recent papers \cite{Jo1} deal with a similar subject:
 beside a detailed study on phase transition in nuclei in the 
framework of the interacting boson model, a summary of different 
solutions of the Bohr hamiltonian is done and an interesting solution for a 
triaxial case with $\gamma \sim \pi/6$ is discussed. The author assumes
a potential of the form $V(\beta,\gamma)= u(\beta)+{1\over 2 }
D\beta^6\cos{3\gamma}$ and follows an approach similar to the one used in
ref. \cite{Iac2}.

The Schr\"odinger equation with the Bohr hamiltonian may be written as
\be
H_B \Psi(\beta,\gamma, \theta_i)\=E  \Psi(\beta,\gamma, \theta_i)
\ee
where $H_B\= T_\beta+T_\gamma+T_{rot}+V(\beta,\gamma)$ \cite{Jean} and the 
kinetic terms have the following expressions
\begin{eqnarray}
T_\beta &\= - \displaystyle { \hbar^2\over 2B_m} 
{1\over \beta^4}{\partial \over 
\partial \beta}\beta^4{\partial \over \partial \beta} \\
T_\gamma &\= -\displaystyle{\hbar^2\over 2B_m} {1\over \beta^2}
 {1\over \sin{(3\gamma)}} {\partial \over \partial \gamma}
\sin{(3\gamma)}{\partial \over \partial \gamma} \\
T_{rot} &\= \displaystyle {\hbar^2\over 8B_m \beta^2}
\sum_{\kappa=1,2,3} {{\hat Q}_\kappa^2 \over 
\sin^2(\gamma-2\pi\kappa/3)} 
\end{eqnarray}
where the ${\hat Q}_\kappa$ are the projections 
of the angular momentum ($\hat L$) 
in the intrinsic frame. The potential term is in general a function of 
$\beta$ and $\gamma$. The conditions, that must be imposed on this 
potential term in 
order to separate the above differential equation, have been discussed 
extensively in the literature (see for example \cite{Bohr,WJ,DavCh}).
For our purposes it will be sufficient to notice that whenever the potential
is chosen as 
\be
V(\beta, \gamma)\= V_1(\beta)+{V_2(\gamma)\over \beta^2}\,,
\label{pote}
\ee
the Schr\"odinger equation is separable \cite{WJ} into the following 
set of second order differential equations:
\begin{eqnarray}
\Bigl(T_\beta+V_1(\beta)-E+{\Omega\over \beta^2}\Bigr) f(\beta) & \=0 \\
\Bigl(\beta^2\bigl(T_\gamma+T_{rot}\bigr) +V_2(\gamma)-\Omega \Bigr) 
\Phi(\gamma,\theta_i) & \=0 
\end{eqnarray}
with  $\Psi(\beta,\gamma, \theta_i) \= f(\beta) \Phi(\gamma,\theta_i)$.
Notice that the second equation does not depend on $\beta$. 
This separation of variables invokes a constant, $\Omega$.
Here we solve the problem of a $\beta-$soft, $\gamma-$soft triaxial 
rotor with a potential that has a minimum located in 
$\beta=\beta_0$ and $\gamma=\gamma_0 \= \pi/6$. 
The simplest approach to this problem is to choose a displaced harmonic 
dependence for the $\gamma$ variable in order to have a pocket around the 
minimum. 
The potential in the asymmetry variable must be an even function of $\gamma$
because of the indistinguishability with respect to the change of sign of 
$\gamma$, and must also be a periodic function to assure that the operation of 
relabeling of the axes does not change the shape of the ellipsoid, but only 
its orientation. This statement is obviously not fulfilled by our 
choice that is to be considered therefore only locally in the region around 
the minimum. We notice that a discrete rotation of $n\pi/3$ in
the polar coordinate system $(\beta,\gamma)$ does not affect our conclusions.

In the $\gamma-$rigid case with a situation of maximum asymmetry (
for $\gamma_0\=\pi/6$) two of the moments of inertia are 'accidentally' equal
\cite{Brink,priv}, while the axes of the ellipsoid have not equal length.
The hamiltonian is thus axially symmetric around
the intrinsic 1-axis, that may be taken as a quantization axis, as pointed out
in ref. \cite{Mey}.
We will restrict to a small region around 
$\pi/6$ to take advantage of this simplification.  Before
discussing how to solve the $\gamma-$soft case, we multiply
the two equations above by $2B_m/\hbar^2$ and we define reduced energy and 
potentials as $\varepsilon\=2B_mE/\hbar^2$ and $u_i(z)\=2B_m V_i(z)/\hbar^2$
with $(i=1,2)$.
Lower case symbols denote reduced quantities and operators.
The new set of 'reduced' equations, given the definitions 
$t_i=(2B_m/\hbar^2)\beta^2\,T_i$ ($i={\gamma,rot}$) and 
$\omega\= 2B_m\Omega/\hbar^2$, is
\begin{eqnarray}
&\Bigl(-\displaystyle{1\over \beta^4}{\partial \over 
\partial \beta}\beta^4{\partial \over \partial \beta}+u_1(\beta)-
\varepsilon+{\omega\over \beta^2}\Bigr) f(\beta) \=0 \label{beta}\\
&\Bigl( t_\gamma+t_{rot} +u_2(\gamma)-\omega \Bigr) 
\Phi(\gamma,\theta_i) \=0 \label{gammang}
\end{eqnarray}
The spectrum is determined by the solution of the first differential
equation in which $\omega$, that it is found from the solution of the 
second differential equation, plays the role of the coefficient
of a 'centrifugal' term and, as we will show, yields a 
non-trivial expression for the energy levels.

Around $\pi/6$ we may set 
$\gamma \= \pi/6+x$, with $x$ a small
expansion parameter. It is easily seen that in this case the 
rotational part of the Bohr hamiltonian, through use of 
trigonometric relations, becomes
$$
\sum_{\kappa=1,2,3} {{\hat Q}_\kappa^2 \over \sin^2(\gamma-2\pi\kappa/3)} \=$$
\be
\= 4(\underbrace{{\hat Q}_1^2+{\hat Q}_2^2+{\hat Q}_3^2)}_{\displaystyle 
{\hat L}^2}+ {\hat Q}_1^2\Biggl({1\over  \cos^2(x)}-4\Biggr)
\label{sum}
\ee 
where we have added and subtracted $4{\hat Q}_1^2$.

The first step to solve the $(\gamma,\theta_i)$ part of the problem is to
change variable in Eq. (\ref{gammang}), using $\gamma= \pi/6+x$ and introducing
the following simplifications:
$$\sin{3\gamma}=\cos{3x}\sim 1 \qquad \,\,\cos{x}\sim 1. $$

We obtain, specifying the harmonic dependence of the potential, 
$u_2(x)\= Cx^2$,
 and exploiting relation (\ref{sum}), the following equation
\be
\Biggl(
\underbrace{ -{\partial^2 \over \partial x^2}+Cx^2-\omega}_{\displaystyle
\hat H_1}+
{\hat{\vec Q}}^2-{3\over 4}{\hat{Q}_1}^2 \Biggr) 
\Phi(x,\theta_i)\=0
\label{demonio}
\ee
where the second order differential operator $\hat H_1$ has been
defined. The wavefunction $\Phi(x,\theta_i)$ may be written, following
\cite{Mey}, as
\be
\sqrt{2L+1 \over 16\pi^2(1+\delta_{R,0})} 
\eta_{n_\gamma,L,R}(x)\bigl[{\cal D}^{(L)}_{M,R}(\theta_i)+(-1)^L
{\cal D}^{(L)}_{M,-R}(\theta_i)\bigr]
\ee
where the angular part is written in terms of Wigner functions labeled 
by the projection of the total angular momentum on the 1-axis, $R$, 
that is a good quantum number, while the functions $\eta_{n_\gamma,L,R}(x)$ 
are eigenfunctions of the one dimensional Schr\"odinger equation for the 
harmonic oscillator, ${\hat H_1}\eta_{n_\gamma,L,R}(x)=0$. The index 
$n_\gamma$ is the quantum number associated with the vibrations in the 
$\gamma$ degree of freedom.
The action of the operators in eq. (\ref{demonio}) on the respective 
wavefunctions gives
\be
\omega_{L,R,n_\gamma} = \sqrt{C}(2n_\gamma+1) + L(L+1)-{3\over 4}R^2 
\label{omega}
\ee
The first term corresponds to a harmonic $\gamma-$vibration, while
the second and third account for the rotational energy of the 
$\gamma=\pi/6$ rigid triaxial rotor, reproducing the Meyer-ter-vehn 
formula \cite{Mey}.

The complete spectrum may then be obtained solving eq. (\ref{beta}) 
with some appropriate potential, $u_1(\beta)$, as, for instance, 
in Ref. \cite{Iac,Iac2,Bohr,WJ,RoBa,FV, Elli,FV2}.
The eigenvalue equation in the $\beta$ variable may in general be written as
\be
\chi''(\beta)+\Bigl( \varepsilon -u_1(\beta) -{\omega+2\over \beta^2}\Bigr)
\chi(\beta) =0
\label{equa}
\ee
having imposed $f(\beta)\=\beta^{-2} \chi(\beta)$. 
When the potential $u_1(\beta)$ is taken to be of a \cite{FV,FV2}
Coulomb-like form $-A/\beta$, or of a Kratzer-like form $-A/\beta+B/\beta^2$,
with the substitutions
$z=2\beta\sqrt{\epsilon}$, $\epsilon=-\varepsilon$, $k=A/(2\sqrt{\epsilon})$
 and $\mu^2=9/4+B+\omega$, we can recast equation (\ref{equa}) as the 
Whittaker's differential equation:
\be
\Bigl\{{d^2\over dz^2} -{1\over 4} +{k\over z}+{1/4-\mu^2\over z^2} \Bigr\}
\chi(z)=0
\ee
whose well-known solution is expressed in terms of Whittaker's functions.
As in \cite{FV,FV2} the requirement of the correct asymptotic behaviour  
fixes the spectrum, introducing the $n_\beta$ quantum number associated 
with $\beta-$vibrations. The reduced eigenvalues are
\be
\epsilon(n_\gamma,n_\beta,L,R)={A^2/4 \over \Bigl(\sqrt{9/4+B+
\omega_{L,R,n_\gamma}} +1/2+n_\beta \Bigr)^2}
\label{spec1}
\ee
where $\omega$ comes from eq. (\ref{omega}). 
Energies are usually redefined fixing the ground state to zero 
and using the energy of the first $2^+$ state as unit, namely
\be
\bar \epsilon(n_\gamma,n_\beta,L,R) = {\epsilon(n_\gamma,n_\beta,L,R)-
\epsilon(0,0,0,0)\over \epsilon(0,0,2,2)-
\epsilon(0,0,0,0)} ~~. 
\ee

\begin{figure}[!t]
\epsfig{file=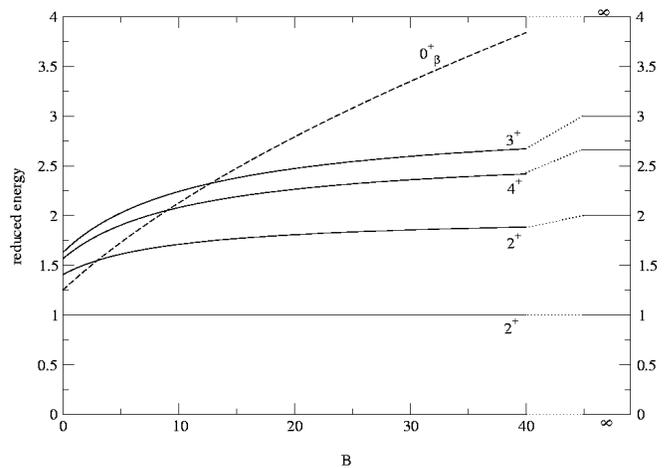,width=0.48\textwidth}
\caption{Reduced energies of the lowest state of the $\beta-$band 
(dashed line) and of a few lowest states of the ground state band 
(solid lines) as a function of $B$. The limits for the energy levels 
when $B\rightarrow \infty$, that correspond to the rigid triaxial rotor
energies, are reported in the right side. Here we fixed $C=1$.}
\label{f1}
\end{figure}
\begin{figure}[!t]\vspace{10mm}
\epsfig{file=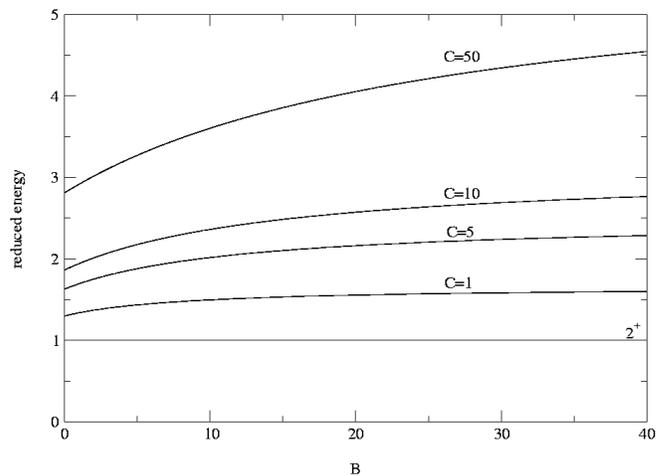,width=0.48\textwidth}
\caption{Reduced energies of the ($J^\pi=2^+,n_\gamma=1,n_\beta=0$) state
(solid line) as a function of $B$, for various values of the strength 
of the harmonic potential in $\gamma$, $C$.
The first $2^+$ of the ground state band (dashed line) is reported for
reference.}
\label{f2}
\end{figure}

In fig. (\ref{f1})
we displayed, as a function of $B$ and for a fixed value of $C=1$, 
the lowest states of the ground state 
band ($n_\beta=0,n_\gamma=0$) and the lowest $0^+$ of the $\beta-$band 
($n_\beta=1,n_\gamma=0$). It may be seen that when $B\rightarrow \infty$
the typical rigid rotor energies are recovered from the limit of the 
ground state energy levels. The relative position of the $\beta-$band 
(eventually tending to infinity), with respect to the ground state band,
gives an idea of the degree of $\beta-$softness.
The negative anharmonicities evidenced in this figure (that are strictly
valid for a triaxial nucleus with $\gamma\sim 30^o$) are qualitatively
consistent with most experimental data on triaxial nuclei where typically 
the $3^+$ and $4^+$ states are found at an energy that is lower than the 
rigid rotor prediction.  
In fig. (\ref{f2}) we plotted the lowest $2^+$ state of the ground state 
band for reference and the $(J^\pi=2^+,n_\gamma=1,n_\beta=0)$ state 
for various values of the parameter $C$, that is the strength
of the harmonic oscillator in $\gamma$. A low value for $C$ corresponds to
a soft well, while a large value corresponds to a more rigid situation.
When $B\rightarrow \infty$ the states of the $\gamma-$band tend to
finite values, as far as $C$ is finite. The Wilets-Jean limit 
(complying with the choice of Coulomb and Kratzer potentials in $\beta$ 
\cite{FV}) is correctly obtained from formula (\ref{spec1}) when the 
$\gamma-$potential becomes flat ($C=0$).
The reduced energy spectra shown in both figures do not depend on the
parameter $A$.
This model has, from one side, a limited applicability because it is 
confined to some special triaxial nuclei for which the minimum of the 
potential lies at $\gamma=\pi/6$, but from the other side, it has 
a considerable flexibility as far as the position of the excited bands
is concerned. The interplay between the two parameters may be exploited
to fit experimental data. Notice that $B$ and $C$ do not separately
fix the position of the respective $\beta$ and $\gamma$ bands, but they 
both take part in a non-trivial way to determine the energy levels. 
It is nevertheless very well known \cite{Cast} that many predictions obtained
for a $\gamma-$rigid model nearly agree with an axially symmetric 
$\gamma-$soft model whenever the $\gamma_{rms}$ of the latter equals the 
rigid value of the former. This fact may increase the applicability of the 
model. 

The addition of $\gamma$ and $\beta$ vibrations to the rigid
triaxial rotor spectrum (around $\gamma=30^o$) shows interesting
features: the well-accepted na\"{\i}ve idea that the
spectrum of the soft rotor is approximatively the sum of a 
rigid rotor term plus separate vibrational terms in $\beta$ and in $\gamma$
is not necessarily correct. Indeed we have shown that, while the 
solution of the $\gamma-$angular part
of the problem gives a straightforward extension of the rigid rotor formula
in which a simple harmonic term for the $\gamma$ degree of freedom appears,
the full spectrum is rather a more complicated function that essentially
depends on the choice of the potential in $\beta$. Other choices are
possible as, for instance, a displaced harmonic oscillator or a Davidson 
potential \cite{RoBa,Elli} in $\beta$, that lead to exact solutions.
The solution presented here may well serve as a good starting point to 
analyze energy spectra in various mass regions where triaxiality is 
expected to play a relevant role.

I wish to thank F.Iachello for having called my attention to this problem,
I.Hamamoto, M.A.Caprio, K.Heyde and  A.Vitturi for valuable 
discussions and D.J.Rowe for enlightening comments.
I acknowledge financial support from FWO-Vlaanderen and from Universit\`a
di Padova, where this work was started.

\end{document}